\documentstyle{article}

\def\dfrac#1#2{{\displaystyle\frac{#1}{#2}}}

\makeatletter
\def\eqnarray{%
  \stepcounter{equation}%
  \let\@currentlabel=\theequation
  \global\@eqnswtrue
  \global\@eqcnt\z@
  \tabskip\@centering
  \let\\=\@eqncr
  $$\halign to \displaywidth\bgroup\@eqnsel\hskip\@centering
  $\displaystyle\tabskip\z@{##}$&\global\@eqcnt\@ne
  \hfil$\displaystyle{{}##{}}$\hfil
  &\global\@eqcnt\tw@$\displaystyle\tabskip\z@{##}$\hfil
  \tabskip\@centering&\llap{##}\tabskip\z@\cr}
\makeatother

\begin{document}

\title{Two-Step Contribution to \\
Intermediate Energy $(p$,$p')$ and $(p$,$n)$ Reactions }

\author{{Y.~Nakaoka
and M.~Ichimura
}\\
                                  \\
        {\it Institute of Physics,} \\
        {\it Graduate School of Arts and Sciences,} \\
        {\it University of Tokyo, Komaba}}
\date{{8-1, Komaba 3 chome, } \\
      {Meguro-ku, TOKYO }  \\
      {153-8902 JAPAN}}

\maketitle

\begin{abstract}

We calculate the two-step contribution to $(p,p')$ and $(p,n)$ 
reactions at intermediate energy.
We describe the motion of the incident nucleon with a plane wave and 
compare the contribution from two-step processes with that from 
one-step processes. 
To describe the two-step processes, we extend the response functions 
to nondiagonal forms with respect to the momentum transfer 
${\bf q}$. 

We performed a numerical calculation for the cross sections of the 
$^{12}$C, $^{40}$Ca$(p,p')$ scattering and the spin longitudinal and 
spin transverse cross sections of the $^{12}$C, $^{40}$Ca$(p,n)$ 
reactions at 346 MeV and 494 MeV.
We found that the two-step contribution is appreciable in comparison 
with the one-step processes in the higher-energy transfer region 
for the spin longitudinal and the spin transverse $(p,n)$ reactions. 
We also found that the two-step processes give larger contributions to 
the spin transverse $(p,n)$ reaction than to the spin longitudinal 
reaction.
This finding is very encouraging to interpret the discrepancy between 
the DWIA calculation and the experimental results of the spin 
longitudinal and the spin transverse cross sections.
\end{abstract}

\section{Introduction}

In nucleon induced high energy reactions, 
single-step processes have often been considered as the main 
contribution to the cross sections.
When the excitation energy is low, 
it is probable that the projectile nucleon collides with the nucleons in 
the target only once, but 
as the excitation energy becomes higher, the pre-equilibrium processes, 
in which the projectile collides with the nucleons several times, 
are considered to become more effective.

Actually, it has been reported that the single-step process calculation 
underestimates the scattering cross sections, especially 
in the large scattering angle region, \cite{Udagawa,Luo} 
where the two-step and further multi-step scattering are found to 
have a large effect.

The multi-step direct reaction~(MSDR) was actively studied at the end 
of the 1970s.
Feshbach, Kerman, and Koonin~(FKK) \cite{FKK,FKL} developed the 
framework of multi-step reaction theory. 
Tamura, Udagawa, and Lenske~(TUL),\cite{Udagawa} who pointed out a 
problem in FKK theory, replaced the sum over the excited nuclear 
eigenstates with that over 1-particle-1-hole~(1p-1h) states, 
introducing a weight function. 
Another type of formalism was presented by Smith and Wambach,
\cite{Wambach} 
who applied the Glauber approximation to the motion of the projectile 
and analyzed the forward angle scattering.
Kawai et al. \cite{Luo,Watanabe,Weiden,Kawai} 
presented a multi-step formalism with a local-density (or Thomas-Fermi) 
approximation for the target and the semiclassical distorted wave~(SCDW) 
approximation for the projectile and the ejectile. 
They calculated the cross sections of $^{58}$Ni, $^{90} $Zr$(p,px)$ and 
$(p,nx)$ reactions at incident energies lower than 200 MeV and 
determined the scattering angle dependence of the one- and the two-step 
processes. 
They extended their calculation to three-step processes 
\cite{Shinohara} and obtained comparable results with one- and 
two-step processes at large angles.
De Pace et al. \cite{DePace} also used the Glauber 
approximation for the analysis of $(p,n)$ reactions and took account of 
the spin transfer.

Although multi-step reactions have been studied by various groups, as 
mentioned above, 
those with spin dependence have been rarely investigated.
In the spin polarized cross sections of the $(p,n)$ reactions, 
a large discrepancy between the distorted wave impulse approximation
(DWIA) calculations and the experimental results has been reported.

In the $(p,n)$ reactions, the spin longitudinal response function $R_L$ 
and the spin transverse response function $R_T$ are extracted from 
the polarization transfer coefficients.
\cite{McClelland,Chen,Taddeucci,Wakasa}
With these observables, the ratio $R_L/R_T$ is found to be less than 1. 

It has been predicted with the random phase approximation~(RPA), 
however, \cite{Alberico} that $R_L$ is 
enhanced and the peak of its energy spectrum is shifted downwards, 
while $R_T$ is quenched and its peak is shifted upwards.\ 
The ratio $R_L/R_T$ becomes greater than 1 theoretically, 
and this contradicts the experimental results. 

The spin longitudinal cross section $ID_q$ 
is roughly reproduced by DWIA with the RPA correlation 
\cite{Wakasa,Agata} in the lower excitation energy region, 
but in the spin transverse cross section $ID_p$, the estimation amounts 
to only about half of the experimental result in the whole 
excitation energy region (see Figs.~1,2). 
In the region where the cross sections are underestimated, 
the multi-step processes are expected to have some contribution. 

For this reason, we calculate the two-step spin longitudinal and the 
spin transverse cross sections of the $(p,n)$ reactions.
These reactions were already studied by De Pace et al.,
but they applied the Glauber approximation, and they assumed that the 
first step momentum transfer and the second step momentum transfer 
are nearly parallel to the total momentum transfer.
They also assumed that the spin flip occurs only in the first or 
second step and that in the other step the spin-isospin scalar 
transition takes place. 
Further, they neglected the interference between processes with 
different shares of the momentum transfer in the first and the second 
steps.

In this paper, we develop a formalism for two-step reactions within 
the framework of the plane-wave approximation in order to see the 
relative contributions from one- and two-step processes. 
We believe that the path lengths of the incident particles in the target 
nucleus in the one-step and the two-step processes are nearly
equal, and therefore the effects of absorption in the one-step processes
and in the two-step processes become nearly equal.
We treat the spin degree of freedom carefully, taking account of the 
difference between the direction of the total momentum transfer and that 
in each step.
With this formalism, we calculate the unpolarized cross section of the 
$(p,p')$ scattering and the spin longitudinal and spin transverse 
cross sections of the $(p,n)$ reactions at intermediate energy. 
We investigate the spin dependence of the two-step contribution 
and partly explain the contradiction in the $R_L/R_T$ problem. 
In \S\ref{formalism} we present our formalism for one- and two-step 
processes in detail and study some relations with TUL's formalism.
In \S\ref{frmspn} we describe the cross sections with spin degrees of 
freedom.
In \S\ref{numcal} we explain our method for the numerical calculation. 
In \S\ref{result} the results and discussion are given, and 
in \S\ref{conclusion} we summarize this paper and give the conclusion.

\section{Formalism}\label{formalism}
\renewcommand{\theequation}{\arabic{section}.\arabic{equation}}
\setcounter{equation}{0}

We consider nucleon induced inelastic scattering and charge exchange 
reactions in which the target is excited to the continuum region. 
The total Hamiltonian $H$ is
\begin{equation}
H=H_0+V \label{eq:total},
\end{equation}
where $H_0$ is the unperturbed Hamiltonian and $V$ is the residual 
interaction between the projectile and the target. They are specified in 
the projectile-target center of mass~(c.m.) system as
\begin{eqnarray}
H_0&=&H_{\rm p}+U+H_{\rm T}, \\
H_{\rm p}&=&\sqrt{M^2+\hat{{\bf p}}^2}, \hspace{0.4cm}
H_{\rm T}=M_{\rm T}+H_{\rm T}^{\rm int}+
            \frac{\hat{{\bf p}}^2}{2M_{\rm T}}, \\
V&=&\sum_{i=1}^Av_{0i}-U\equiv{\cal V}-U, \\
H_{\rm T}^{\rm int}&=&
  H_{{\rm T}}^{{\rm shell}} + V_{{\rm T}}, \\
H_{{\rm T}}^{{\rm shell}}&=&
  \sum_{i=1}^A \left[ \frac{\hat{{\bf p}}_i^2}{2M}+
  U_i^\prime \right], \hspace{0.4cm}
V_{{\rm T}}=\sum_{i<j}v_{ij} - \sum_{i=1}^AU_i^\prime ,
\end{eqnarray}
where $H_{\rm p}$ and $H_{\rm T}$ are the projectile and the target 
Hamiltonian, respectively, and $v$ is the effective interaction between 
particles. 
The subscript $0$ represents the projectile and $i$ designates the 
$i$-th nucleon in the target.
$U$ is the mean field that the target nucleus creates for the projectile.
$M$ and $M_{\rm T}$ are the nucleon mass and the ground state target 
mass, respectively, 
$\hat{{\bf p}}$ is the projectile momentum operator in the 
projectile-target c.m.~frame, 
and $\hat{{\bf p}}_i$ is the momentum operator of the $i$-th 
nucleon in the intrinsic frame of the target.
The $U_i^\prime$ are the mean fields for the nucleons in the target.

The $T$-matrix is given in the `solved' Lippman-Schwinger form
\begin{equation}
T=V+V\dfrac{1}{E^{+}-H}V ,
\end{equation}
where $E^+=E+i\epsilon$.
Using (\ref{eq:total}) and expanding $1/(E^{+}-H)$ in $V$,
one obtains
\begin{equation}
  T=\sum_{j=1}^{\infty} V(G_0V)^{j-1} ,
\end{equation}
with
\begin{equation}
  G_0=\dfrac{1}{E^{+}-H_0} .
\end{equation}
Then, the $T$-matrix elements become
\begin{equation}
T_{n0}^{{\rm fi}}=\sum_{j=1}^{\infty}T_{n0}^{{\rm fi}(j)} \label{eq:lipp},
\end{equation}
with the $j$-th order $T$-matrices
\begin{equation}
T_{n0}^{{\rm fi}(j)}\equiv
  \langle \chi_{\rm f}^{(-)}|\langle \Phi_n|V(G_0V)^{j-1}|\Phi_0\rangle 
  |\chi_{\rm i}^{(+)} \rangle ,
\end{equation}
where $\chi_{\rm i}^{(+)}$ and $\chi_{\rm f}^{(-)}$ are the distorted 
waves of the initial and final channel, and the eigenstates of the 
target nucleus $|\Phi_n\rangle$ obey the equation
\begin{equation}
H_{\rm T}^{\rm int}|\Phi_n\rangle =E_n^{\rm int}|\Phi_n\rangle .
\end{equation}
Here $n=0$ denotes the ground state, and we set $E_0^{\rm int}=0$.
The $j$-th term of (\ref{eq:lipp}) describes the $j$-step processes.

Here we neglect the interference terms with $j\neq j^\prime$.
The different order $T$-matrices mainly excite different number 
particle-hole states.
Then, the final state of $T^{(j)}$ is different from that of 
$T^{(j^\prime)}$.
Hence the contribution from the interference terms can be neglected.
Then the double-differential cross section becomes
\begin{eqnarray}
\dfrac{\partial^2\sigma}
  {\partial \Omega_{\rm f} \partial \epsilon_{\rm f}}
&\simeq&\sum_{j=1}^\infty
\dfrac{\partial^2\sigma^{(j)}}
  {\partial \Omega_{\rm f} \partial \epsilon_{\rm f}}
=K\sum_{j=1}^\infty X^{(j)}, \\
K&\equiv&\dfrac{\mu_{\rm i}\mu_{\rm f}}{(2\pi)^2}
         \dfrac{k_{\rm f}}{k_{\rm i}} , \hspace{0.4cm}
X^{(j)}\equiv\sum_n|T_{n0}^{{\rm fi}(j)}|^2 
  \delta(\omega-(E_{\rm f}-E_{\rm i})), \label{eq:ij}
\end{eqnarray}
where $\omega~(=\epsilon_{\rm i}-\epsilon_{\rm f})$ is the energy 
transfer, $k_{\rm i}$ and $k_{\rm f}$ are the projectile and ejectile 
asymptotic wave number, 
$\mu_{\rm i}\ (\mu_{\rm f})$ is the reduced energy of the 
projectile~(ejectile),
\begin{equation}
  \mu_{\rm i}=\frac{\epsilon_{\rm i} E_{\rm i}}
                     {\epsilon_{\rm i}+E_{\rm i}},\hspace{0.5cm}
  \mu_{\rm f}=\frac{\epsilon_{\rm f} E_{\rm f}}
                     {\epsilon_{\rm f}+E_{\rm f}}, 
\end{equation}
and $\epsilon_{\rm i}\ (\epsilon_{\rm f})$ and $E_{\rm i}\ (E_{\rm f})$ are 
the projectile~(ejectile) and the target~(residual) nucleus energies,
\begin{equation}
\epsilon_{\rm i}=\sqrt{M^2+k_{\rm i}^2},\hspace{0.5cm}
\epsilon_{\rm f}=\sqrt{M^2+k_{\rm f}^2},
\end{equation}
\begin{equation}
E_{\rm i}=M_{\rm T}+
            \frac{k_{\rm i}^2}{2M_{\rm T}},\hspace{0.5cm}
E_{\rm f}=M_{\rm T}+E_n^{\rm int}+
            \frac{k_{\rm f}^2}{2M_{\rm T}}.
\end{equation}
\subsection{One-step processes}

We are mainly interested in the relative importance of the one- and the 
two-step processes.
Therefore we adopt a plane-wave approximation for 
the distorted waves $\chi_{\rm i}^{(+)}$ and $\chi_{\rm f}^{(-)}$.
Then the one-step $T$-matrix becomes
\begin{equation}
T_{n0}^{(1)}({\bf k}_{\rm f},{\bf k}_{\rm i}) \equiv
  \langle {\bf k}_{\rm f}|\langle \Phi_n|V|
  \Phi_0 \rangle|{\bf k}_{\rm i}\rangle 
=\langle {\bf k}_{\rm f}|\langle \Phi_n|{\cal V}|
  \Phi_0 \rangle|{\bf k}_{\rm i}
  \rangle 
\end{equation}
for $n\neq0$. 
The effective interaction ${\cal V}$ between the projectile and the 
target can be written as
\begin{equation}
{\cal V}({\bf r}_0) \equiv \sum_{i=1}^A v_{0i}({\bf r}_0-{\bf r}_i)
=\int v({\bf r}_0-{\bf r}) \rho ({\bf r})d{\bf r},
\end{equation}
where 
$\rho({\bf r})=\sum_{i=1}^A\delta({\bf r}-{\bf r}_i)$.
For the moment we suppress the spin-isospin dependence of the 
interaction $v$ to make the argument simple.
From \S3.1 we explicitly introduce the spin dependence.

In the impulse approximation the effective interaction 
$v({\bf r}_0-{\bf r})$ between the projectile and the nucleon in the 
target is approximated by
\begin{equation}
v({\bf r}_0-{\bf r})=\int t_{\rm NN}({\bf p})
e^{i{\bf p} \cdot ({\bf r}_0-{\bf r})}\frac{d{\bf p}}{(2\pi)^3},
\end{equation}
where $t_{\rm NN}({\bf p})$ is the nucleon-nucleon transition matrix
(NN $t$-matrix) for the momentum transfer ${\bf p}$.
Then
\begin{equation}
\langle {\bf k}_{\rm f}|{\cal V}({\bf r}_0)|{\bf k}_{\rm i} \rangle 
= t_{\rm NN}({\bf k}_{\rm f}-{\bf k}_{\rm i})
  \rho({\bf k}_{\rm f}-{\bf k}_{\rm i}) . \label{eq:tp}
\end{equation}
Defining the momentum transfer as 
${\bf q}\equiv {\bf k}_{\rm f}-{\bf k}_{\rm i}$, one finds
\begin{equation}
T_{n0}^{(1)}({\bf k}_{\rm f},{\bf k}_{\rm i})=
t_{\rm NN}({\bf q})
\langle \Phi_n|\rho({\bf q})|\Phi_0 \rangle  \label{eq:trho},
\end{equation}
where $\rho({\bf q})$ is the density operator in the momentum space.
This is the $T$-matrix in the $t\rho$ approximation, which is a simple 
version in various impulse approximations.
The NN $t$-matrix is, in fact, 
expressed as a function of ${\bf q}$, 
${\bf Q}(\equiv {\bf k}_{\rm i}+{\bf k}_{\rm f})$, 
and the incident energy in the laboratory frame $K_{\rm lab}$, 
but we suppressed the ${\bf Q}$ dependence, 
and we do not write $K_{\rm lab}$ explicitly.

Substituting (\ref{eq:trho}) into (\ref{eq:ij}), 
one obtains $X^{(1)}$ as
\begin{eqnarray}
X^{(1)}&=&|t_{\rm NN}({\bf q})|^2  \sum_{n\neq0}
   |\langle \Phi_n|\rho({\bf q})|\Phi_0 \rangle |^2 
   \delta(\omega-(E_{\rm f}-E_{\rm i})) ,\\
&=&\frac{\sqrt{s}}{M_{\rm R}}
   |t_{\rm NN}({\bf q})|^2 R({\bf q},\omega^{\rm int}) 
   \label{eq:Smith} ,
\end{eqnarray}
where $R({\bf q},\omega^{\rm int})$ is the response function of the 
density fluctuation 
\begin{equation}
R({\bf q},\omega^{\rm int})\equiv
   \sum_{n\neq0}\langle \Phi_0|\rho^\dagger({\bf q})|\Phi_n\rangle
             \langle \Phi_n|\rho({\bf q})|\Phi_0\rangle
             \delta(\omega^{\rm int}-E_n^{\rm int}). 
             \label{eq:response}
\end{equation}
The quantity $M_{\rm R}~(=M_{\rm T}\,+\,E_n^{\rm int})$ is the mass of the 
residual nucleus, and $s=(\epsilon_{\rm i}\,+\,E_{\rm i})^2$.
The response function defined in (\ref{eq:response}) is the strength 
per unit energy with respect to the energy transfer in the intrinsic 
frame $\omega^{\rm int}$, while the double-differential cross 
section is the cross section per unit energy with respect to the 
energy transfer in the projectile-target c.m.~system 
$\omega$.
The factor $|d\omega^{\rm int}/d\omega|=\sqrt{s}/M_{\rm R}$ is the 
variable transformation coefficient derived from the equation 
\cite{Hagedorn}
\begin{equation}
\epsilon_{\rm f}=\frac{s+M^2-M_{\rm R}^2}{2\sqrt{s}} .
\end{equation}
Thus the one-step double-differential cross section can be written as 
a product of the square of the NN $t$-matrix and the response function.

Since the exact nuclear states $|\Phi_n\rangle$ are complicated, 
we cannot calculate the response functions by using the expression 
(\ref{eq:response}).
To obtain a calculable expression for the response functions, we adopt 
two approximations of TUL.\cite{Udagawa}
They assumed that $|\Phi_0\rangle $ is the 0-particle-0-hole~(0p-0h) 
state and divided the excited state $|\Phi_n\rangle$ into 1p-1h 
states $|\Phi_B\rangle $ and other more complicated states and wrote it 
as
\begin{equation}
|\Phi_n\rangle =\sum_{B} a_{B}^{(n)}|\Phi_B\rangle + \delta|\Phi_n
   \rangle ,
 \ \ \  n\neq0 .
\end{equation}
With the equation
\begin{equation}
\langle \Phi_n|\rho({\bf q})|\Phi_0\rangle =
   \sum_Ba_B^{(n)*}\langle \Phi_B|\rho({\bf q})|\Phi_0\rangle , 
\end{equation}
which implies that the density operator $\rho({\bf q})$ can excite the 
target only to 1p-1h states, 
(\ref{eq:response}) becomes 
\begin{equation}
R({\bf q},\omega^{\rm int})=
   \sum_{n\neq0}\sum_{B^\prime B}
             \langle \Phi_0|\rho^\dagger({\bf q})|\Phi_B\rangle
             a_{B^\prime}^{(n)}a_B^{(n)*}
             \langle \Phi_B|\rho({\bf q})|\Phi_0\rangle
             \delta(\omega^{\rm int}-E_n^{\rm int}). 
\end{equation}
Then we use the TUL statistical approximation
\begin{equation}
\sum_{n\neq0}a_{B}^{(n)*}a_{B^\prime}^{(n)}
   \delta(\omega^{\rm int}-E_n^{\rm int})=
   \delta_{BB^\prime}c_{B}(\omega^{\rm int})  \label{eq:strength},
\end{equation}
by assuming that the $a_B^{(n)}$ are random with respect to $n$ for a 
fixed $B$.
For $B=B^\prime$, the left-hand side of (\ref{eq:strength}) becomes 
$c_B(\omega^{\rm int})$, the probability per unit energy that the state 
$|\Phi_B\rangle$ is mixed in at the excitation energy 
$\omega^{\rm int}$. 
Then we obtain the relation
\begin{equation}
R({\bf q},\omega^{\rm int})\simeq
\sum_{B}c_B(\omega^{\rm int})
\langle \Phi_0|\rho^\dagger({\bf q})|\Phi_B \rangle
\langle \Phi_B|\rho({\bf q})|\Phi_0 \rangle  \label{eq:corr} .
\end{equation}
This is calculable with a reasonable assumption for 
$c_B(\omega^{\rm int})$.
\subsection{Two-step processes}\label{second}

The two-step $T$-matrix in the plane wave approximation can be written as
\begin{equation}
T_{n0}^{(2)}({\bf k}_{\rm f},{\bf k}_{\rm i})
=\sum_{n^\prime\neq0}
\langle {\bf k}_{\rm f}|\langle \Phi_n|{\cal V}|
  \Phi_{n^\prime}\rangle G_{n^\prime}
\langle \Phi_{n^\prime}|{\cal V}|\Phi_0\rangle |{\bf k}_{\rm i}\rangle 
  \label{eq:tsecond} ,
\end{equation}
where the propagator $G_{n^\prime}$ is
\begin{equation}
G_{n^\prime}=\left\langle \Phi_{n^\prime}\left|\frac{1}{E^+-H_0}
  \right|\Phi_{n^\prime} \right\rangle .
\end{equation}
We rewrite (\ref{eq:tsecond}) as
\begin{equation}
T_{n0}^{(2)}({\bf k}_{\rm f},{\bf k}_{\rm i})=
\sum_{n^\prime\neq0}\int
\dfrac{d{\bf k}_{\rm m}^\prime }{(2\pi)^3}
\dfrac{d{\bf k}_{\rm m}}{(2\pi)^3}
\langle {\bf k}_{\rm f}|\langle \Phi_n|{\cal V}|\Phi_{n^\prime} \rangle
|{\bf k}_{\rm m}^\prime \rangle
\langle {\bf k}_{\rm m}^\prime|G_{n^\prime}|{\bf k}_{\rm m} \rangle 
\langle {\bf k}_{\rm m} |\langle \Phi_{n^\prime}|{\cal V}|
  \Phi_0\rangle|{\bf k}_{\rm i}\rangle .
\end{equation}
Since we are interested in high incident energy reactions, 
we can assume that the mean field $U$ has a small effect on the 
particle.
We assume that the momentum of the propagating particle does 
not vary during the sequential collisions, and thus
\begin{equation}
U=U_0({\bf k}_{\rm m})\delta({\bf k}_{\rm m}-{\bf k}_{\rm m}^\prime).
\end{equation}
The Green's function becomes diagonal with respect to 
${\bf k}_{\rm m}$, so we can write the $T$-matrix as
\begin{equation}
T_{n0}^{(2)}({\bf k}_{\rm f},{\bf k}_{\rm i})
=\sum_{n^\prime\neq0}\int \dfrac{d{\bf k}_{\rm m}}{(2\pi)^3} 
\langle{\bf k}_{\rm f}|\langle \Phi_n|{\cal V}|\Phi_{n^\prime} \rangle
|{\bf k}_{\rm m} \rangle
G_{n^\prime}({\bf k}_{\rm m})
\langle{\bf k}_{\rm m}|\langle \Phi_{n^\prime}|{\cal V}|
  \Phi_0\rangle|{\bf k}_{\rm i}\rangle , \label{eq:T2}
\end{equation}
where 
\begin{equation}
G_{n^\prime}({\bf k}_{\rm m})\equiv\dfrac{1}
  {E^+ -U_0({\bf k}_{\rm m}) -\sqrt{M^2+{\bf k}_{\rm m}^2} 
  -\left(M_{\rm T}+E_{n^\prime}^{\rm int}
  +\dfrac{{\bf k}_{\rm m}^2}{2M_{\rm T}} \right)  } .
\end{equation}

Now we use the two TUL assumptions.
The first one is
\begin{equation}
\langle \Phi_{n^\prime}|{\cal V}|\Phi_0\rangle =
\sum_Ca_C^{(n^\prime)*}\langle \Phi_C|{\cal V}|\Phi_0\rangle 
  \label{eq:ph}.
\end{equation}
The other is the `spectator assumption',
\begin{equation}
\langle \Phi_n|{\cal V}|\Phi_{n^\prime} \rangle
  =\sum_Ba_{Bn^\prime}^{(n)*}
  \langle \Phi_B|{\cal V}|\Phi_0\rangle \label{eq:spectator},
\end{equation}
which implies that a collision always occurs by creating a new 
particle-hole pair and that the particle-hole pairs which are present 
in $|\Phi_{n^\prime} \rangle$ play only the role of spectators.
Substituting (\ref{eq:ph}), (\ref{eq:spectator}) and (\ref{eq:tp}) into 
(\ref{eq:T2}) we obtain 
\begin{eqnarray}
T_{n0}^{(2)}({\bf k}_{\rm f},{\bf k}_{\rm i})
&=&\sum_{n^\prime\neq0}\sum_{BC}
  \int \dfrac{d{\bf q}_1}{(2\pi)^3}
  a_{Bn^\prime}^{(n)*} 
  t_{\rm NN}({\bf q}-{\bf q}_1)
  \langle \Phi_B|\rho({\bf q}-{\bf q}_1)|\Phi_0\rangle  \nonumber \\
& &\ \ \ \ \ \times
  G_{n^\prime}({\bf k}_{\rm i}+{\bf q}_1) 
  a_C^{(n^\prime)*} t_{\rm NN}({\bf q}_1)
  \langle \Phi_C|\rho({\bf q}_1)|\Phi_0\rangle ,
\end{eqnarray}
where ${\bf q}_1={\bf k}_{\rm m}-{\bf k}_{\rm i}$.

Next we proceed to calculate $X^{(2)}$ of (\ref{eq:ij}).
It becomes 
\begin{eqnarray}
X^{(2)}&=&\frac{\sqrt{s}}{M_{\rm R}}
  \sum_{n\neq0}\delta(\omega^{\rm int}-E_n^{\rm int})
  \sum_{n^\prime\neq0}
  \sum_{n^{\prime\prime}\neq0}\sum_{BC}\sum_{B^\prime C^\prime}
  \int \dfrac{d{\bf q}_1}{(2\pi)^3}
  \int \dfrac{d{\bf q}_1^\prime}{(2\pi)^3}
  a_{Bn^\prime}^{(n)*}a_{B^\prime n^{\prime\prime}}^{(n)} 
  a_C^{(n^\prime)*}a_{C^\prime}^{(n^{\prime\prime})}  \nonumber \\
& \times& t_{\rm NN}({\bf q}-{\bf q}_1)
  t_{\rm NN}^\dagger({\bf q}-{\bf q}_1^\prime)
  \langle \Phi_0|\rho^\dagger({\bf q}-{\bf q}_1^\prime)
  |\Phi_{B^\prime}\rangle
  \langle \Phi_B|\rho({\bf q}-{\bf q}_1)|\Phi_0\rangle \nonumber \\
& \times& G_{n^{\prime\prime}}^{*}({\bf k}_{\rm i}+{\bf q}_1^\prime)
  G_{n^\prime}({\bf k}_{\rm i}+{\bf q}_1)
  t_{\rm NN}({\bf q}_1)
  t_{\rm NN}^\dagger({\bf q}_1^\prime)
  \langle \Phi_0|\rho^\dagger({\bf q}_1^\prime)|\Phi_{C^\prime}\rangle
  \langle \Phi_C|\rho({\bf q}_1)|\Phi_0 \rangle . \nonumber \\
& &
\end{eqnarray}
Substituting the TUL assumption 
\begin{equation}
\sum_{n\neq0}\delta(\omega^{\rm int}-E_n^{\rm int})
a_{Bn^\prime}^{(n)*}a_{B^\prime n^{\prime\prime}}^{(n)}
=\delta_{BB^\prime}\delta_{n^\prime n^{\prime\prime}}
 c_{Bn^\prime}(\omega^{\rm int}) ,
\end{equation}
which is analogous to (\ref{eq:strength}), we obtain
\begin{eqnarray}
X^{(2)}&=&\frac{\sqrt{s}}{M_{\rm R}}\sum_{n^\prime\neq0}
  \sum_{BCC^\prime}\int \dfrac{d{\bf q}_1}{(2\pi)^3}
  \int \dfrac{d{\bf q}_1^\prime}{(2\pi)^3} \nonumber \\
&\times&
  c_{Bn^\prime}(\omega^{\rm int})
  a_C^{(n^\prime)*}a_{C^\prime}^{(n^\prime)}
  t_{\rm NN}({\bf q}-{\bf q}_1)
  t_{\rm NN}^\dagger({\bf q}-{\bf q}_1^\prime) \nonumber  \\
&\times&\langle \Phi_0|\rho^\dagger({\bf q}-{\bf q}_1^\prime)
  |\Phi_B\rangle
  \langle \Phi_B|\rho({\bf q}-{\bf q}_1)|\Phi_0\rangle 
  G_{n^\prime}({\bf k}_{\rm i}+{\bf q}_1)
  G_{n^\prime}^{*}({\bf k}_{\rm i}+{\bf q}_1^\prime) \nonumber \\
&\times& t_{\rm NN}({\bf q}_1)
  t_{\rm NN}^\dagger({\bf q}_1^\prime)
  \langle \Phi_0|\rho^\dagger({\bf q}_1^\prime)|\Phi_{C^\prime}\rangle
  \langle \Phi_C|\rho({\bf q}_1)|\Phi_0 \rangle .
\end{eqnarray}
With the identity 
$\displaystyle \int d\omega_1
  \delta(\omega_1-(E_{\rm m}-E_{\rm i}))=1$, 
$X^{(2)}$ becomes
\begin{eqnarray}
X^{(2)}&=&\frac{\sqrt{s}}{M_{\rm R}}
  \sum_{n^\prime\neq0}\sum_{BCC^\prime}
  \int d\omega_1
  \int \dfrac{d{\bf q}_1}{(2\pi)^3}
  \int \dfrac{d{\bf q}_1^\prime}{(2\pi)^3} \nonumber \\
& \times&
  \delta(\omega_1-(E_{\rm m}-E_{\rm i}))
  c_{Bn^\prime}(\omega^{\rm int})
  a_C^{(n^\prime)*}a_{C^\prime}^{(n^\prime)}    \nonumber  \\
& \times& t_{\rm NN}({\bf q}-{\bf q}_1)
  t_{\rm NN}^\dagger({\bf q}-{\bf q}_1^\prime)
  \langle \Phi_0|\rho^\dagger({\bf q}-{\bf q}_1^\prime)|\Phi_B\rangle
  \langle \Phi_B|\rho({\bf q}-{\bf q}_1)|\Phi_0 \rangle 
  \nonumber \\
&\times& G({\bf k}_{\rm i}+{\bf q}_1,\omega_1)
  G^{*}({\bf k}_{\rm i}+{\bf q}_1^\prime,\omega_1)
  t_{\rm NN}({\bf q}_1)
  t_{\rm NN}^\dagger({\bf q}_1^\prime)
  \langle \Phi_0|\rho^\dagger({\bf q}_1^\prime)|\Phi_{C^\prime} \rangle
  \langle \Phi_C|\rho({\bf q}_1)|\Phi_0 \rangle , \nonumber \\
& & \label{eq:x2delta} 
\end{eqnarray}
where $E_{\rm m}$ is the target energy in the intermediate state,
\begin{equation}
E_{\rm m}=
  M_{\rm T}+E_{n^\prime}^{\rm int}
  +\frac{{\bf k}_{\rm m}^2}{2M_{\rm T}}.
\end{equation}
Here we have eliminated the $n^\prime$ dependence of 
$G_{n^\prime}({\bf k}_{\rm i}+{\bf q}_1)$ as
\begin{equation}
G({\bf k}_{\rm i}+{\bf q}_1,\omega_1)=
  \dfrac{1}
  {E^+ -U_0({\bf k}_{\rm m}) -\sqrt{M^2+{\bf k}_{\rm m}^2} 
  -\left(M_{\rm T}+
        \dfrac{{\bf k}_{\rm i}^2}{2M_{\rm T}}+
        \omega_1 \right) } .
\end{equation}
Rewriting the delta function with $\omega_1^{\rm int}$, 
(\ref{eq:x2delta}) becomes
\begin{eqnarray}
X^{(2)}&=&\frac{s}{M_{\rm R}^2}
  \sum_{n^\prime\neq0}\sum_{BCC^\prime}
  \int d\omega_1
  \int \dfrac{d{\bf q}_1}{(2\pi)^3}
  \int \dfrac{d{\bf q}_1^\prime}{(2\pi)^3}
  \delta(\omega_1^{\rm int}-E_{n^\prime}^{\rm int})
  c_{Bn^\prime}(\omega^{\rm int})
  a_C^{(n^\prime)*}a_{C^\prime}^{(n^\prime)}    \nonumber  \\
&\times& t_{\rm NN}({\bf q}-{\bf q}_1)
  t_{\rm NN}^\dagger({\bf q}-{\bf q}_1^\prime)
  \langle \Phi_0|\rho^\dagger({\bf q}-{\bf q}_1^\prime)|\Phi_B\rangle
  \langle \Phi_B|\rho({\bf q}-{\bf q}_1)|\Phi_0 \rangle \nonumber \\
&\times& G({\bf k}_{\rm i}+{\bf q}_1,\omega_1)
  G^{*}({\bf k}_{\rm i}+{\bf q}_1^\prime,\omega_1)
  t_{\rm NN}({\bf q}_1)
  t_{\rm NN}^\dagger({\bf q}_1^\prime)
  \langle \Phi_0|\rho^\dagger({\bf q}_1^\prime)|\Phi_{C^\prime} \rangle
  \langle \Phi_C|\rho({\bf q}_1)|\Phi_0 \rangle .\nonumber \\
& &
\end{eqnarray}
Using the approximation 
\begin{equation}
c_{Bn^\prime}(\omega^{\rm int})\simeq 
  c_B(\omega^{\rm int}-\omega_1^{\rm int}),
\end{equation}
which results from the assumption (\ref{eq:spectator}) that the 
particle-hole pairs which are present in the intermediate state have 
nothing to do with creating a new particle-hole pair,
we apply the TUL approximation (\ref{eq:strength}) as
\begin{eqnarray}
\hspace*{-1cm}X^{(2)}&=&\frac{s}{M_{\rm R}^2}\int d\omega_1
  \int \dfrac{d{\bf q}_1}{(2\pi)^3}
  \int \dfrac{d{\bf q}_1^\prime}{(2\pi)^3} \nonumber \\
&\times&
  t_{\rm NN}({\bf q}-{\bf q}_1)
  t_{\rm NN}^\dagger({\bf q}-{\bf q}_1^\prime) \nonumber \\
&\times&
  \sum_B c_B (\omega^{\rm int}-\omega_1^{\rm int})
  \langle \Phi_0|\rho^\dagger({\bf q}-{\bf q}_1^\prime)|\Phi_B\rangle
  \langle \Phi_B|\rho({\bf q}-{\bf q}_1)|\Phi_0 \rangle
        \nonumber \\
&\times&
  G({\bf k}_{\rm i}+{\bf q}_1;\omega_1)
  G^{*}({\bf k}_{\rm i}+{\bf q}_1^\prime;\omega_1) \nonumber \\
&\times&
  t_{\rm NN}({\bf q}_1)
  t_{\rm NN}^\dagger({\bf q}_1^\prime)
  \sum_Cc_C(\omega_1^{\rm int})
  \langle \Phi_0|\rho^\dagger({\bf q}_1^\prime)|\Phi_C\rangle
  \langle \Phi_C|\rho({\bf q}_1)|\Phi_0 \rangle .
\end{eqnarray}
Consequently, we obtain
\begin{eqnarray}
\hspace*{-1cm}X^{(2)}&=&\frac{s}{M_{\rm R}^2}\int d\omega_1
  \int \dfrac{d{\bf q}_1}{(2\pi)^3}
  \int \dfrac{d{\bf q}_1^\prime}{(2\pi)^3} \nonumber \\
&\times&
  t_{\rm NN}({\bf q}-{\bf q}_1)
  t_{\rm NN}^\dagger({\bf q}-{\bf q}_1^\prime)
  R({\bf q}-{\bf q}_1,{\bf q}-{\bf q}_1^\prime;
    \omega^{\rm int}-\omega_1^{\rm int}) \nonumber \\
&\times&
  G({\bf k}_{\rm i}+{\bf q}_1;\omega_1)
  G^{*}({\bf k}_{\rm i}+{\bf q}_1^\prime;\omega_1)
  t_{\rm NN}({\bf q}_1)
  t_{\rm NN}^\dagger({\bf q}_1^\prime)
  R({\bf q}_1,{\bf q}_1^\prime;\omega_1^{\rm int}) \label{eq:i2},
\end{eqnarray}
where $R({\bf q},{\bf q}^\prime;\omega^{\rm int})$ is expressed as
\begin{equation}
R({\bf q},{\bf q}^\prime;\omega^{\rm int})=
\displaystyle\sum_Ac_A(\omega^{\rm int})
\langle \Phi_0|\rho^\dagger({\bf q}^\prime)|\Phi_A\rangle
\langle \Phi_A|\rho({\bf q})|\Phi_0\rangle .
\end{equation}
Here we have extended the response function to a nondiagonal form with 
respect to the momentum transfer ${\bf q}$.
The last three factors in (\ref{eq:i2}), which is the product of two NN 
$t$-matrices and the response function, represent the first collision, 
and with the next product of the Green's functions, the particle in the 
intermediate state propagates in the target nucleus. 
Then the second collision occurs with the remaining three factors.

\section{Formalism with spin}\label{frmspn}
\setcounter{equation}{0}

We are concerned with the unpolarized cross section of the $(p,p')$ 
scatterings and the spin longitudinal and spin transverse cross 
sections of the $(p,n)$ reactions.
Hence we must treat the spin of the nucleon explicitly.
First we give the form of the unpolarized cross section and define 
\cite{Kawahigashi} the spin longitudinal and spin transverse 
cross sections.
We introduce the unit vectors
\begin{equation}
\hat{\bf q}=\frac{\bf q}{|\bf q|}, \ \ \
\hat{\bf n}=\frac{{\bf k}_{\rm i}\times{\bf k}_{\rm f}}
                 {|{\bf k}_{\rm i}\times{\bf k}_{\rm f}|}, \ \ \
\hat{\bf p}=\hat{\bf q}\times\hat{\bf n}.
\end{equation}
The scattering $T$-matrix is generally written as 
\begin{equation}
T({\bf k}_{\rm f},{\bf k}_{\rm i})=
  \hat{T}_0+\hat{T}_n\sigma_0\cdot\hat{n}+
  \hat{T}_q\sigma_0\cdot\hat{q}+\hat{T}_p\sigma_0\cdot\hat{p} , 
\end{equation}
where $\sigma_0$ is the spin operator of the incident nucleon and 
the $\hat{T}_i$ are the operators of the target.
In this section we suppress the isospin degree of freedom.

The unpolarized differential cross section $I$ is 
given by
\begin{equation}
I=\frac{1}{2(2J_{\rm T}+1)}K{\rm Tr}{\rm Tr}^\prime [TT^\dagger],
\end{equation}
with the target angular momentum $J_{\rm T}$.
The symbol Tr represents the trace of the spin substates of the 
scattering nucleon and Tr$^\prime$ denotes the summation of 
all allowed initial and final states of the target,
\begin{equation}
{\rm Tr}^\prime[TT^\dagger]=\sum_{0,n}
  \langle \Phi_0|T^\dagger|\Phi_n\rangle \langle \Phi_n|T|\Phi_0\rangle 
  \delta(\omega-(E_{\rm f}-E_{\rm i})).
\end{equation}

The observables $D_i$ introduced by Bleszynski et al. 
\cite{Bleszynski} are
\begin{eqnarray}
ID_0&=&\frac{1}{2J_{\rm T}+1}K{\rm Tr}^\prime
      [\hat{T}_0\hat{T}_0^\dagger], \label{eq:id0} \\
ID_n&=&\frac{1}{2J_{\rm T}+1}K{\rm Tr}^\prime
      [\hat{T}_n\hat{T}_n^\dagger], \label{eq:idn} \\
ID_q&=&\frac{1}{2J_{\rm T}+1}K{\rm Tr}^\prime
      [\hat{T}_q\hat{T}_q^\dagger], \label{eq:idq} \\
ID_p&=&\frac{1}{2J_{\rm T}+1}K{\rm Tr}^\prime
      [\hat{T}_p\hat{T}_p^\dagger]. \label{eq:idp}
\end{eqnarray}
We refer to $ID_q$ and $ID_p$ as the `spin longitudinal' and `spin 
transverse' cross sections respectively.
\cite{McClelland,Taddeucci}
In the following, we only consider the case $J_{\rm T}=0$, 
because we assume that $\Phi_0$ is the 0p-0h state.
\subsection{One-step processes}\label{onestepspin}

The NN $t$-matrices can be decomposed generally as
\begin{equation}
t_{\rm NN}({\bf q})=\sum_{\mu\bar{\mu}}
   \sigma_{0\mu}\sigma_{i\bar{\mu}}t_{\mu\bar{\mu}}({\bf q}), 
   \label{eq:tnngen}
\end{equation}
with $\mu,\bar{\mu}=u,q,n,p$, 
where $\sigma_{ju}=I_j,\sigma_{j\mu}=\sigma_j\cdot\hat{\mu}\ 
(\mu\neq u), (j=0,i)$.
The operator $\sigma_i$ denotes the spin operator of the nucleon in the 
target.
With this equation, (\ref{eq:Smith}) can be rewritten as
\begin{eqnarray}
X^{(1)}&=&\sum_{\mu \mu^\prime}
      X_{\mu^\prime \mu}^{(1)}\sigma_{0\mu^\prime}\sigma_{0\mu}, \\
X_{\mu^\prime \mu}^{(1)}&\equiv&\frac{\sqrt{s}}{M_{\rm R}}
   \sum_{\bar{\mu}\bar{\mu}^\prime}
   t_{\mu^\prime \bar{\mu}^\prime}^*({\bf q})
   R_{\bar{\mu}^\prime \bar{\mu}}({\bf q},\omega^{\rm int})
   t_{\mu\bar{\mu}}({\bf q}), 
\end{eqnarray}
where
\begin{eqnarray}
R_{\bar{\mu}^\prime \bar{\mu}}({\bf q},\omega^{\rm int})&\equiv&
   \sum_{B}c_B(\omega^{\rm int})
   \langle \Phi_0|\rho_{\bar{\mu}^\prime}^\dagger({\bf q})
   |\Phi_B \rangle
   \langle \Phi_B|\rho_{\bar{\mu}}({\bf q})|\Phi_0 \rangle , \\
\rho_{\bar{\mu}}({\bf q})&\equiv&
   \sum_{i=1}^A
   e^{-i{\bf q}\cdot{\bf r}_i} 
   \sigma_{i\bar{\mu}} .
\end{eqnarray}

The unpolarized, the spin longitudinal and the spin transverse cross 
sections, $I$, $ID_q$ and $ID_p$, can be rewritten as 
\begin{equation}
I   = K\sum_{\mu}X_{\mu\mu}^{(1)}, \hspace{0.5cm}
ID_q= KX_{qq}^{(1)}, \hspace{0.5cm}
ID_p= KX_{pp}^{(1)}, \label{eq:fstspn}
\end{equation}
respectively.
\subsection{Two-step processes}

In the spin dependent two-step formalism, 
$X^{(2)}$ in (\ref{eq:i2}) becomes 
\begin{eqnarray}
X^{(2)}&=&\int d\omega_1
      \int \dfrac{d{\bf q}_1}{(2\pi)^3}
      \int \dfrac{d{\bf q}_1^\prime}{(2\pi)^3} \nonumber \\
&\times&
      \sum_{\mu_2\mu_2^\prime}\sum_{\mu_1\mu_1^\prime}
      X_{\mu_2^\prime \mu_2\mu_1^\prime \mu_1}^{(2)}
      \sigma_{0\mu_1^\prime}\sigma_{0\mu_2^\prime}
      \sigma_{0\mu_2}\sigma_{0\mu_1}, \label{eq:x2} \\
X_{\mu_2^\prime \mu_2\mu_1^\prime \mu_1}^{(2)}&\equiv&
  \frac{s}{M_{\rm R}^2}
  \sum_{\bar{\mu}_2\bar{\mu}_2^\prime}
  \sum_{\bar{\mu}_1\bar{\mu}_1^\prime}
  t_{\mu_1^\prime\bar{\mu}_1^\prime}^*({\bf q}_1^\prime)
  t_{\mu_2^\prime\bar{\mu}_2^\prime}^*({\bf q}-{\bf q}_1^\prime)
  t_{\mu_2\bar{\mu}_2}({\bf q}-{\bf q}_1)
  t_{\mu_1\bar{\mu}_1}({\bf q}_1) \nonumber \\
&\times&
  G({\bf k}_{\rm i}+{\bf q}_1;\omega_1)
  G^{*}({\bf k}_{\rm i}+{\bf q}_1^\prime;\omega_1) \nonumber \\
&\times&
  R_{\bar{\mu}_2^\prime \bar{\mu}_2}
    ({\bf q}-{\bf q}_1,{\bf q}-{\bf q}_1^\prime;
    \omega^{\rm int}-\omega_1^{\rm int})
  R_{\bar{\mu}_1^\prime \bar{\mu}_1}
    ({\bf q}_1,{\bf q}_1^\prime;\omega_1^{\rm int}),\label{eq:x2mu}
\end{eqnarray}
where
$\mu_1,\bar{\mu_1}=u,q_1,n_1,p_1$, and $\mu_2,\bar{\mu_2}=u,q_2,n_2,p_2$,
 and so on.
Here we defined the unit vectors
\begin{equation}
\hat{\bf q}_2=\frac{{\bf q}-{\bf q}_1}{|{\bf q}-{\bf q}_1|}, \ \ \
\hat{\bf n}_2=\frac{{\bf k}_{\rm f}\times\hat{\bf q}_2}
                  {|{\bf k}_{\rm f}\times\hat{\bf q}_2|}, \ \ \
\hat{\bf p}_2=\hat{\bf q}_2\times\hat{\bf n}_2.
\end{equation}
We also extended the response function to nondiagonal form 
with respect to the spin direction as
\begin{equation}
R_{\bar{\mu}^\prime \bar{\mu}}
  ({\bf q},{\bf q}^\prime;\omega^{\rm int})=
\sum_A c_A(\omega^{\rm int})
\langle \Phi_0|\rho_{\bar{\mu}^\prime}^\dagger({\bf q}^\prime)
  |\Phi_A\rangle
\langle \Phi_A|\rho_{\bar{\mu}}({\bf q})|\Phi_0\rangle .
\end{equation}
The unpolarized, the spin longitudinal, and the spin transverse cross 
sections become 
\begin{eqnarray}
I&=&K\frac{1}{2}{\rm Tr}X^{(2)}=
    K\int d\omega_1
    \int \dfrac{d{\bf q}_1}{(2\pi)^3}
    \int \dfrac{d{\bf q}_1^\prime}{(2\pi)^3} \nonumber \\
&\times&
    \sum_{\mu_2\mu_2^\prime}\sum_{\mu_1\mu_1^\prime}
    X_{\mu_2^\prime \mu_2\mu_1^\prime \mu_1}^{(2)}
    \frac{1}{2}{\rm Tr}
   (\sigma_{0\mu_1^\prime}\sigma_{0\mu_2^\prime}
    \sigma_{0\mu_2}\sigma_{0\mu_1}), \label{eq:sndI}\\
ID_q&=&K
    \int d\omega_1
    \int \dfrac{d{\bf q}_1}{(2\pi)^3}
    \int \dfrac{d{\bf q}_1^\prime}{(2\pi)^3} \nonumber \\
&\times&
    \sum_{\mu_2\mu_2^\prime}\sum_{\mu_1\mu_1^\prime}
    X_{\mu_2^\prime \mu_2\mu_1^\prime \mu_1}^{(2)}
    \frac{1}{2}{\rm Tr}
    (\sigma_{0\mu_1^\prime}\sigma_{0\mu_2^\prime}\sigma_q)
    \frac{1}{2}{\rm Tr}
    (\sigma_q\sigma_{0\mu_2}\sigma_{0\mu_1}), \label{eq:idq2} \\
ID_p&=&K
    \int d\omega_1
    \int \dfrac{d{\bf q}_1}{(2\pi)^3}
    \int \dfrac{d{\bf q}_1^\prime}{(2\pi)^3} \nonumber \\
&\times&
    \sum_{\mu_2\mu_2^\prime}\sum_{\mu_1\mu_1^\prime}
    X_{\mu_2^\prime \mu_2\mu_1^\prime \mu_1}^{(2)}
    \frac{1}{2}{\rm Tr}
    (\sigma_{0\mu_1^\prime}\sigma_{0\mu_2^\prime}\sigma_p)
    \frac{1}{2}{\rm Tr}
    (\sigma_p\sigma_{0\mu_2}\sigma_{0\mu_1}), \label{eq:sndIDp}
\end{eqnarray}
respectively.

\section{Practical method of calculation}\label{numcal}
\subsection{Various processes of two-step reactions}\label{pattern}
\setcounter{equation}{0}
 
We calculate $(p,p')$ inelastic scattering and $(p,n)$ charge exchange 
reactions at 346 and 494 MeV. \ \
In the two-step $(p,n)$ reaction, the $(p,n)$ charge exchange collision 
occurs in the first step or in the second step (see Fig.~3).
If the $(p,n)$ charge exchange collision occurs in the first step, the 
second step has to be a $(n,n')$ collision, and if the first step is 
a $(p,p')$ collision, then the second step has to be a $(p,n)$ 
collision. 

Moreover, there are two cases in non-charge-exchange $(p,p')$ and 
$(n,n')$ collisions, because the nucleon in the target nucleus that 
participates in the collisions can be a proton or a neutron. 
Consequently, we must consider four cases in the two-step $(p,n)$ 
reaction.

For two-step $(p,p')$ scattering, we must consider five 
cases (see Fig.~4). 
If the $(p,n)$ charge exchange collision occurs in the first step, then 
the second step has to be a $(n,p)$ collision. 
In this case, the nucleon in the target nucleus that takes part in the 
first step collision has to be a neutron and that in the second step 
collision has to be a proton. 
This is the only case in which charge exchange collisions occur in 
the two-step $(p,p')$ scattering. 

However, if no charge exchange collision occurs, the nucleons struck in 
the first step and the second step can be a proton or a neutron. 
Hence we can consider four cases in the two-step $(p,p')$ scattering 
that include no charge exchange collisions. 
\subsection{Green's function}

Our aim is to compare the two-step contribution with the one-step 
contribution.
We adopted the plane wave approximation for the motion of the projectile 
and the ejectile.
Thus in the one-step processes no absorption was included.
We believe that the path lengths of the incident particles in the target 
nucleus in the one-step and in the two-step processes are nearly equal, 
and the effects of absorption become nealy equal in these 
processes.
Therefore we should also remove the effect of absorption from the 
Green's function.
Then the effect of absorption in the two-step processes become zero, as 
in the one-step processes.
We use the relativistic expression for the Green's function 
and apply the on-energy shell approximation as 
\begin{equation}
G({\bf k}_{\rm m})\simeq -i\pi
  \delta\left(E -U_0 -\sqrt{M^2+{\bf k}_{\rm m}^2} 
  -\sqrt{(M_{\rm T}+\omega_1^{\rm int})^2
  +{\bf k}_{\rm m}^2}\right) 
  \label{eq:imaginary} .
\end{equation}
We take a representative value for ${\bf k}_{\rm m}$ in $U_0$ and 
set $U_0=-$3.5 MeV\cite{Mottelson} and 5.5 MeV\cite{Jones} for 
$K_{\rm lab}=346$ MeV and 494 MeV, respectively. 

Substituting (\ref{eq:imaginary}) into (\ref{eq:x2mu}), we obtain, 
for instance, from (\ref{eq:sndI})
\begin{eqnarray}
I&=&K\int_0^\omega d\omega_1 
  \int_0^\infty \dfrac{q_1^2}{(2\pi)^3}dq_1
  \int_{-1}^1 d({\rm cos}\theta_1)
  \int_0^{2\pi} d\phi_1
  \int_0^\infty \dfrac{q_1^{\prime 2}}{(2\pi)^3}dq_1^\prime
  \int_{-1}^1 d({\rm cos}\theta_1^\prime)
  \int_0^{2\pi} d\phi_1^\prime  \nonumber \\
&\times&
  \sum_{\mu_2\mu_2^\prime}\sum_{\mu_1\mu_1^\prime}
  X_{\mu_2^\prime \mu_2\mu_1^\prime \mu_1}^{(2)}
  \frac{1}{2}{\rm Tr}
  (\sigma_{0\mu_1^\prime}\sigma_{0\mu_2^\prime}
   \sigma_{0\mu_2}\sigma_{0\mu_1}), \label{eq:x2shell}
\end{eqnarray}
where
\begin{eqnarray}
X_{\mu_2^\prime\mu_2\mu_1^\prime\mu_1}^{(2)}
  &=&\frac{\pi^2s}{M_{\rm R}^2}
  \sum_{\bar{\mu}_2\bar{\mu}_2^\prime}
  \sum_{\bar{\mu}_1\bar{\mu}_1^\prime}
  \nonumber \\
&\times&
  t_{\mu_1^\prime\bar{\mu}_1^\prime}^*(K_{\rm lab},{\bf q}_1^\prime)
  t_{\mu_2^\prime\bar{\mu}_2^\prime}^*
    (K_{\rm lab}^\prime,{\bf q}-{\bf q}_1^\prime)
  t_{\mu_2\bar{\mu}_2}(K_{\rm lab}^\prime,{\bf q}-{\bf q}_1)
  t_{\mu_1\bar{\mu}_1}(K_{\rm lab},{\bf q}_1) \nonumber \\
&\times&
  \delta\left(E-U_0
  -\sqrt{M^2+({\bf k}_{\rm i}+{\bf q}_1)^2} 
  -\sqrt{(M_{\rm T}+\omega_1^{\rm int})^2
  +({\bf k}_{\rm i}+{\bf q}_1)^2}\right) \nonumber \\
&\times&
  \delta\left(E-U_0 -\sqrt{M^2
             +({\bf k}_{\rm i}+{\bf q}_1^\prime)^2}
   -\sqrt{(M_{\rm T}+\omega_1^{\rm int})^2+
      ({\bf k}_{\rm i}+{\bf q}_1^\prime)^2}\right) \nonumber \\
&\times&
  R_{\bar{\mu}_2 \bar{\mu}_2^\prime}
    ({\bf q}-{\bf q}_1,{\bf q}-{\bf q}_1^\prime;
    \omega^{\rm int}-\omega_1^{\rm int})
  R_{\bar{\mu}_1\bar{\mu}_1^\prime}
    ({\bf q}_1,{\bf q}_1^\prime;\omega_1^{\rm int}). 
    \label{eq:x2shellmu}
\end{eqnarray}
We set $z$-axis parallel to the direction of ${\bf k}_{\rm i}$. 
With the delta functions, we can carry out the integration of 
${\rm cos}\theta_1$ and ${\rm cos}\theta_1^\prime$ analytically. 

Other integrations are carried out numerically.
The mesh sizes were chosen as 
$\Delta\omega_1^{\rm int}=10.0$ MeV, $\Delta q_1^{\rm int}=0.2$~(1/fm) 
and $\Delta\phi_1=15.0^\circ$, where $q_1^{\rm int}=\{(A-1)/A\} q_1$, 
with which we can obtain sufficiently accurate results.
The integration range of $q_1^{\rm int}$ is restricted up to 
$4.4$~(1/fm), beyond which the response functions are negligibly small.

Here we have written the energy dependence of the NN $t$-matrices 
explicitly and assumed that the projectile has kinetic energy 
$K_{\rm lab}$ in the first collision and $K_{\rm lab}^\prime$ in the 
second collision, where 
\begin{equation}
K_{\rm lab}^\prime=K_{\rm lab}-\omega_{1{\rm lab}}, \hspace{0.4cm}
\omega_{1{\rm lab}}=
  \frac{-t-M_{\rm T}^2+M_{\rm R}^2}{2M_{\rm T}}, \hspace{0.4cm}
t=\omega_1^2-q_1^2 .
\end{equation}
\subsection{NN $t$-matrices}\label{nntmatrix}

We use the method of Love and Franey \cite{Love} 
to calculate the NN $t$-matrix.
The derivations of the formulae are given in detail in the Appendix. 
We use the parameter values of Ref.~\cite{Franey} and treat the 
exchange terms by replacing $Q_{\rm cm}$ with $k_{\rm i}$, as is 
suggested in Ref.~\cite{Love}.

The scattering amplitude $f_{\rm NN}(K_{\rm lab},q_{\rm cm})$ 
is written as \cite{Kerman}
\begin{eqnarray}
f_{\rm NN}(K_{\rm lab},q_{\rm cm})&=&
    (A_0+A_1(\tau_0\cdot\tau_i))
  + (B_0+B_1(\tau_0\cdot\tau_i))
    \sigma_0\cdot\hat{\bf n}_{\rm cm} 
    \sigma_i\cdot\hat{\bf n}_{\rm cm}  \nonumber \\
&+&(C_0+C_1(\tau_0\cdot\tau_i))
     \sigma_0 \cdot\hat{\bf n}_{\rm cm}
  + (C_0+C_1(\tau_0\cdot\tau_i))
     \sigma_i \cdot\hat{\bf n}_{\rm cm}  \nonumber  \\
&+&(E_0+E_1(\tau_0\cdot\tau_i))
   \sigma_0\cdot\hat{\bf q}_{\rm cm} \sigma_i\cdot\hat{\bf q}_{\rm cm}
   \nonumber \\
 &+&(F_0+F_1(\tau_0\cdot\tau_i))
   \sigma_0\cdot\hat{\bf Q}_{\rm cm} \sigma_i\cdot\hat{\bf Q}_{\rm cm},
   \label{eq:sixterm}
\end{eqnarray}
where 
\begin{equation}
\hat{\bf q}_{\rm cm}=
  \frac{{\bf k}_{\rm cm}^\prime-{\bf k}_{\rm cm}}
       {|{\bf k}_{\rm cm}^\prime-{\bf k}_{\rm cm}|}, \ \ \
\hat{\bf n}_{\rm cm}=
  \frac{{\bf k}_{\rm cm}\times{\bf k}_{\rm cm}^\prime}
       {|{\bf k}_{\rm cm}\times{\bf k}_{\rm cm}^\prime|}, \ \ \
\hat{\bf Q}_{\rm cm}=\hat{\bf q}_{\rm cm}\times\hat{\bf n}_{\rm cm}, 
\end{equation}
and ${\bf k}_{\rm cm}\ ({\bf k}_{\rm cm}^\prime)$ is the projectile
(ejectile) momentum in the NN c.m.~system. 

The relation between the NN $t$-matrices and the NN scattering 
amplitudes in the NN c.m.~frame is
\begin{eqnarray}
t_{\rm NN}(K_{\rm lab},q_{\rm cm})&=&
  \eta f_{\rm NN}(K_{\rm lab},q_{\rm cm}), \label{eq:ATOT} \\
\eta&=&\frac{-4\pi}{\sqrt{M^2+k_{\rm cm}^2}}.
\end{eqnarray}
Therefore the $t_{\mu\bar{\mu}}$ in (\ref{eq:tnngen}) are given by
\begin{eqnarray}
t_{uu}&=&\eta(A_0+A_1(\tau_0\cdot\tau_i)), \\
t_{nn}&=&\eta(B_0+B_1(\tau_0\cdot\tau_i)), \\
t_{nu}&=&\eta(C_0+C_1(\tau_0\cdot\tau_i)), \\
t_{un}&=&\eta(C_0+C_1(\tau_0\cdot\tau_i)), \\
t_{qq}&=&\eta(E_0+E_1(\tau_0\cdot\tau_i)), \\
t_{pp}&=&\eta(F_0+F_1(\tau_0\cdot\tau_i)),
\end{eqnarray}
and all others are 0.

Calculating the expectation values of the isospin operators, 
we get the amplitudes for proton-proton scattering as
\begin{equation}
t_{uu}=\eta(A_0+A_1) ,\cdots .
\end{equation}
For neutron-neutron scattering these are the same as for proton-proton 
scattering.

Other types of NN collisions are proton-neutron charge exchange 
collisions and proton-neutron non-charge exchange collisions.
Their amplitudes are
\begin{eqnarray}
t_{uu}&=&2\eta A_1, \cdots , \label{eq:pnchex}\\
t_{uu}&=&\eta(A_0-A_1), \cdots ,
\end{eqnarray}
respectively.

We note that the isospin operators of the struck nucleons are included 
in the response functions. Hence we use the $t$-matrix amplitudes 
for the charge exchange collision as 
\begin{equation}
t_{uu}=\sqrt{2}\eta A_1, \cdots ,
\end{equation}
instead of (\ref{eq:pnchex}).
\subsection{Response functions}\label{response function}

We obtained a simple expression for the response function in 
(\ref{eq:corr}).
We take account of the weight function $c_B(\omega)$ 
through a complex potential for the single particles, 
which is of a Woods-Saxon-type central potential with the derivative 
form spin-orbit potential as
\begin{eqnarray}
U^\prime(r)&=&
  -(V+iW)\dfrac{1}{1+\exp\left(\dfrac{r-R_0}{a_0}\right)} \nonumber \\
&-&2\left(\frac{1}{m_\pi}\right)^2
  \frac{V_{ls}}{a_{\rm SO}}
  \dfrac{\exp\left(\dfrac{r-R_{\rm SO}}{a_{\rm SO}}\right)}
    {r\left[1+\exp\left(\dfrac{r-R_{\rm SO}}{a_{\rm SO}}\right)\right]^2}
   ({\bf l}\cdot{\bf s})
  +V_{\rm Coul},
\end{eqnarray}
where the radius parameters are $r_0=r_{\rm SO}=r_{\rm Coul}=1.27$ fm, 
the diffusenesses are $a_0=a_{\rm SO}=0.67$ fm, 
the potential depths are determined by the binding energy of the last 
occupied single particle level, and the imaginary part of the potential 
is $W=5.0$ MeV. \cite{Gaarde}
The depth of the ${\bf l}\cdot{\bf s}$ part is chosen to be 6.5 MeV
(10.0 MeV) for $^{12}$C ($^{40}$Ca). \cite{Mottelson}
The response functions used in the DWIA calculation depicted in Figs.~%
1 and 2 include the effects of the RPA correlation with the $\Delta$ 
degree of freedom as well as the effective mass and the spreading width 
of hole states.
However, here we calculate the response functions with no correlations 
in order to extract the effect of the nuclear reaction mechanism 
exclusively.

We extended the response functions to nondiagonal form with respect 
to the momentum transfer ${\bf q}$ in the two-step formalism. 
The spin scalar isoscalar response function in the nondiagonal form is 
defined as \cite{Nishida}
\begin{equation}
R({\bf q},{\bf q}^\prime;\omega^{\rm int})\equiv\sum_{B\neq0}
  \left \langle \Phi_0\left|\sum_i e^{i{\bf q}^\prime \cdot{\bf r}_i}
  \right |\Phi_B\right\rangle
  \left \langle \Phi_B\left|\sum_j e^{-i{\bf q}\cdot{\bf r}_j}
  \right |\Phi_0\right\rangle
   \delta(\omega^{\rm int}-E_B^{\rm int}). \label{eq:ndrsppp}
\end{equation}

The response functions which include the spin operators are generally 
written in the form 
\begin{eqnarray}
R_{\hat{\bf u}^\prime \hat{\bf u}}
   ({\bf q},{\bf q}^\prime;\omega^{\rm int})&&\nonumber\\
  && \hspace*{-1cm}\equiv\sum_{B\neq0}
  \left \langle \Phi_0\left|\sum_i \sigma_i\cdot\hat{\bf u}^\prime
                  e^{i{\bf q}^\prime \cdot{\bf r}_i}\right|\Phi_B\right\rangle
  \left \langle \Phi_B\left|\sum_j \sigma_j\cdot\hat{\bf u}
                  e^{-i{\bf q}\cdot{\bf r}_j}\right|\Phi_0\right\rangle
   \delta(\omega^{\rm int}-E_B^{\rm int}).\nonumber\\
\end{eqnarray}
Here the response function has also been extended to the nondiagonal 
form with respect to the spin direction.

If the spin-orbit force is negligible and the target nucleus is 
spin-saturated, the single particle states can be written as 
$|n,l,m,s=1/2,\mu \rangle$ with the principal quantum number $n$, 
the angular momentum $l$, its third component $m$, the spin $s$, 
and its third component $\mu$.
Then the response function can be separated into the spin part and the 
orbital part as
\begin{eqnarray}
R_{\hat{\bf u}^\prime \hat{\bf u}}
   ({\bf q},{\bf q}^\prime;\omega^{\rm int})
   &\equiv&\sum_{n_pl_pm_p}\sum_{n_hl_hm_h}\sum_{\mu_p\mu_h}
   \left[\langle \mu_h|\sigma\cdot\hat{\bf u}^\prime|\mu_p\rangle 
         \langle \mu_p|\sigma\cdot\hat{\bf u}|\mu_h\rangle \right] 
   \nonumber \\
&\times&
   \left[
   \langle n_hl_hm_h|e^{i{\bf q}^\prime \cdot{\bf r}}|n_pl_pm_p\rangle
   \langle n_pl_pm_p|e^{-i{\bf q}\cdot{\bf r}}|n_hl_hm_h\rangle \right]
   \delta(\omega^{\rm int}-E_B^{\rm int}), \nonumber \\
& &
\end{eqnarray}
where {\it p} and {\it h} denote a particle and a hole, respectively.
The spin part becomes 
Tr$[\sigma\cdot\hat{\bf u}^\prime \sigma\cdot\hat{\bf u}]
  =2\hat{\bf u}^\prime \cdot \hat{\bf u}$, 
while that of the response function without the spin operators becomes 
the trace of the unit matrix. 
Thus we get the relation
\begin{equation}
R_{\hat{\bf u}^\prime \hat{\bf u}}
({\bf q},{\bf q}^\prime;\omega^{\rm int})=
(\hat{\bf u}^\prime \cdot \hat{\bf u})
R({\bf q},{\bf q}^\prime;\omega^{\rm int}). \label{eq:spnrsp}
\end{equation}
This relation is plotted in Fig.~5, and we judge it 
reasonable.

If only one spin operator is included in the response function, 
then the spin part becomes Tr$[\sigma\cdot\hat{\bf u}]=0$, 
and the response functions become
\begin{eqnarray}
R_{\hat{\bf u}}
   ({\bf q},{\bf q}^\prime;\omega^{\rm int})
   &\equiv&\sum_{B\neq0}
  \left \langle \Phi_0\left|\sum_i e^{i{\bf q}^\prime \cdot{\bf r}_i}
                 \right|\Phi_B\right\rangle
  \left \langle \Phi_B\left|\sum_j \sigma_j\cdot\hat{\bf u}
                  e^{-i{\bf q}\cdot{\bf r}_j}\right|\Phi_0\right\rangle
   \delta(\omega^{\rm int}-E_B^{\rm int})\nonumber\\
&=&0, \nonumber \\
R_{\hat{\bf u}^\prime}
   ({\bf q},{\bf q}^\prime;\omega^{\rm int})
   &\equiv&\sum_{B\neq0}
  \left \langle \Phi_0\left|\sum_i \sigma_i\cdot\hat{\bf u}^\prime
                  e^{i{\bf q}^\prime \cdot{\bf r}_i}\right|\Phi_B\right\rangle
   \left\langle \Phi_B\left|\sum_j e^{-i{\bf q}\cdot{\bf r}_j}
                 \right|\Phi_0\right\rangle
   \delta(\omega^{\rm int}-E_B^{\rm int})\nonumber\\
&=&0. \nonumber \\
   & & \label{eq:onespin}
\end{eqnarray}

We cannot calculate the response functions for charge exchange 
collisions with these relations. 
Therefore we must prepare two more response functions.
One is for the $(p,n)$ collision and the other is for the $(n,p)$ 
collision.
The difference among these three response functions is seen in Fig.~6.
These response functions also satisfy the relations
\begin{eqnarray}
R_{\hat{\bf u}^\prime \hat{\bf u}}^{+}
({\bf q},{\bf q}^\prime;\omega^{\rm int})&=&
(\hat{\bf u}^\prime \cdot \hat{\bf u})
R^{+}({\bf q},{\bf q}^\prime;\omega^{\rm int}), \\
R_{\hat{\bf u}^\prime \hat{\bf u}}^{-}
({\bf q},{\bf q}^\prime;\omega^{\rm int})&=&
(\hat{\bf u}^\prime \cdot \hat{\bf u})
R^{-}({\bf q},{\bf q}^\prime;\omega^{\rm int}),
\end{eqnarray}
where
\begin{eqnarray}
R^{+}({\bf q},{\bf q}^\prime;\omega^{\rm int})&\equiv&\sum_{B\neq0}
   \left\langle \Phi_0\left|\sum_i e^{i{\bf q}^\prime \cdot{\bf r}_i}\tau_-
                 \right|\Phi_B\right\rangle
   \left\langle \Phi_B\left|\sum_j e^{-i{\bf q}\cdot{\bf r}_j}\tau_+
                 \right|\Phi_0\right\rangle
   \delta(\omega^{\rm int}-E_B^{\rm int}), \nonumber \\
R^{-}({\bf q},{\bf q}^\prime;\omega^{\rm int})&\equiv&\sum_{B\neq0}
   \left\langle \Phi_0\left|\sum_i e^{i{\bf q}^\prime \cdot{\bf r}_i}\tau_+
                 \right|\Phi_B\right\rangle
   \left\langle \Phi_B\left|\sum_j e^{-i{\bf q}\cdot{\bf r}_j}\tau_-
                 \right|\Phi_0\right\rangle
   \delta(\omega^{\rm int}-E_B^{\rm int}), \nonumber \\
& &
\end{eqnarray}
with $\tau_\pm=(\tau_x\pm i\tau_y)/\sqrt{2}$.
The response functions $R^+$ and $R^-$ are for the $(p,n)$ and $(n,p)$ 
collisions, respectively.
With these three response functions, it is sufficient to calculate 
the various response functions that appear in (\ref{eq:x2shellmu}).
Using the `spectator assumption', we assume that the target is in the 
ground state of $^{12}$C ($^{40}$Ca), 
even when $(n,n')$ or $(n,p)$ collision occurs in the second step.

We must consider the struck nucleons when we use the response 
function for $(p,p')$ collisions.
If we write the response function for $(p,p')$ collisions as 
(\ref{eq:ndrsppp}), 
the struck nucleon can be a proton or a neutron.
However, when we calculate the cross sections,
we distinguish a proton from a neutron, as shown in Figs.~3 and 4.
Therefore we must divide the spin scalar isoscalar response function 
by 2 when we apply this response function to (\ref{eq:x2shellmu}).
\subsection{Cross sections}
\label{spnexp}

Using the spin-dependent form of the cross sections in 
(\ref{eq:fstspn}), 
the one-step unpolarized, the spin longitudinal and transverse cross 
sections are written as
\begin{eqnarray}
I 
&=&K\frac{\sqrt{s}}{M_{\rm R}}
   \sum_{\mu\bar{\mu}}
   |t_{\mu\bar{\mu}}({\bf q})|^2
   R({\bf q},\omega^{\rm int}), \\
ID_q
&=&K\frac{\sqrt{s}}{M_{\rm R}}
   |t_{qq}({\bf q})|^2
   R_{qq}({\bf q},\omega^{\rm int}) , \\
ID_p
&=&K\frac{\sqrt{s}}{M_{\rm R}}
   |t_{pp}({\bf q})|^2
   R_{pp}({\bf q},\omega^{\rm int}), 
\end{eqnarray}
respectively.
Here we have used the approximation 
${\bf q} \simeq {\bf q}_{\rm cm}$ which was justified in 
Ref.~\cite{Kawahigashi}.
The summations of $\mu$ and $\bar{\mu}$ run over only $(\mu,\bar{\mu})=
(u,u)$, $(n,n)$, $(n,u)$, $(u,n)$, $(q,q)$ and $(p,p)$, 
and we have used the relation 
$R_{\bar{\mu}\bar{\mu}}({\bf q},\omega^{\rm int})
 =R({\bf q},\omega^{\rm int})$, 
which is derived from (\ref{eq:spnrsp}).

We now explain the method of calculating the two-step cross sections
given in (\ref{eq:sndI})--(\ref{eq:sndIDp}).
The quantity $X_{\mu_2^\prime \mu_2\mu_1^\prime \mu_1}^{(2)}$ has four 
NN $t$-matrices, each of which generates 6 terms.
Therefore we encounter $6^4~(=1296)$ terms for each isospin-dependent 
process explained in \S\ref{pattern}.
However, with the approximation (\ref{eq:onespin}) we can eliminate many 
terms, and finally we have to take only 400 terms 
(see Fig.~7). 
We see that the response functions are common except the scalar 
products that appear from the spin parts (see (\ref{eq:spnrsp}) and so 
on).

Next we carry out the summation of 400 products of four NN $t$-matrices 
with the scalar products.
The traces of the spin operators of the incident nucleon are 
easily calculated as, for instance, 
\begin{eqnarray}
\frac{1}{2}{\rm Tr}(\sigma_q\sigma_{0\mu_2}\sigma_{0\mu_1})
&=&{\bf s}\cdot\hat{\bf q}, \\
\frac{1}{2}{\rm Tr}(\sigma_p\sigma_{0\mu_2}\sigma_{0\mu_1})
&=&{\bf s}\cdot\hat{\bf p},
\end{eqnarray}
with the relation
\begin{equation}
\sigma_{0\mu_2}\sigma_{0\mu_1}=I_0s_0+\sigma_0\cdot{\bf s}, 
\end{equation}
where $s_0$ is a real number and ${\bf s}$ is a complex vector.
After the summation we calculate the five-fold integration of the sum 
multiplied by the spin scalar response functions.

In fact, the spin scalar response functions are also common if charge 
exchange collision occurs in the same step in the $(p,n)$ reactions.
As for the $(p,p')$ scattering, 
they are common to the four processes that have no charge exchange 
collision (see Figs.~3 and 4). 
Hence, in a practical calculation, we also carry out the summation of 
the products of four NN $t$-matrices with the scalar products in these 
processes, in which the spin scalar response functions are common, and 
after the summation we calculate the five-fold integration.

\section{Results}\label{result}

In Fig.~8 we display the double differential cross sections 
of the $(p,p')$ scattering and the spin longitudinal and the spin 
transverse $^{12}$C$(p,n)$ reactions at 494 MeV as functions of the 
energy transfer $\omega_{\rm lab}$ in the laboratory frame. 
In this calculation the momentum transfer $q^{\rm int}$ in the intrinsic 
frame of the target is 1.55~(1/fm) ($q_{\rm lab}$ in the laboratory frame 
is 1.69--1.72~(1/fm)).
The scattering angle corresponding to this momentum transfer is about 
18$^\circ$. \cite{McClelland,Chen,Taddeucci} 

In the $^{12}$C$(p,p')$ scattering, the peak of the one-step cross 
section is around $\omega_{\rm lab}=60$ MeV in this no-correlation 
calculation, and the contribution of the two-step processes there is 
only about 0.5$\%$ of the one-step process.
The two-step cross section is still increasing around $\omega_{\rm lab}
=125$ MeV, and it becomes about 6.0$\%$ of the one-step process there. 
These results can be seen in panel (d), which indicates the ratios of 
the two-step cross sections to the one-step cross sections.
The contribution of the two-step processes to the $^{12}$C$(p,p')$ 
scattering is negligibly small.
However, in the $^{12}$C$(p,n)$ reactions, there are considerable 
contributions from the two-step processes.
In the spin longitudinal $^{12}$C$(p,n)$ reaction, the contribution 
of the two-step processes is 3.5$\%$ of the one-step process at 
$\omega_{\rm lab}=60$ MeV and 28$\%$ at $\omega_{\rm lab}=125$ MeV.
For the spin transverse $^{12}$C$(p,n)$ reaction, the two-step 
contribution becomes larger than that for the spin longitudinal 
reaction, and at $\omega_{\rm lab}=60$ and 125 MeV the ratios become 
9.0$\%$ and 68$\%$, respectively.
From these results we can say that the two-step processes become more 
effective as the energy transfer increases in the quasi-elastic region.

Figure 9 displays the double differential cross section 
at 346 MeV of the same reactions. 
The momentum transfer $q^{\rm int}$ is 1.52~(1/fm) ($q_{\rm lab}$ is 
1.66--1.68~(1/fm)).
It corresponds to a scattering angle of 22$^\circ$.\cite{Wakasa} \
The contributions from the two-step processes become larger than those
at 494 MeV. \   
In the $(p,p')$ scattering, the ratio at $\omega_{\rm lab}=60$ MeV is 
less than 1$\%$ and at $\omega_{\rm lab}=125$ MeV it is about 11$\%$.
In the spin longitudinal $^{12}$C$(p,n)$ reaction, the ratio is close 
to 40$\%$ at $\omega_{\rm lab}=125$ MeV, 
while that of the spin transverse $^{12}$C$(p,n)$ reaction exceeds 
75$\%$ there.
The contribution to the spin transverse reaction is much larger than 
that to the spin longitudinal reaction, as in the case of 494 MeV.

We assume that the $E$-term and the $F$-term in the NN $t$-matrices 
mainly determine the spin longitudinal and the spin transverse cross 
sections, respectively, in the two-step processes.
The difference between the two-step contributions for the spin 
longitudinal and the spin transverse cross sections may 
be due to the momentum transfer dependence of the $E$-term and the 
$F$-term which are shown in Fig.~10.
The dominant real part of the $E$-term changes sign at 0.7~(1/fm), 
while the $F$-term never becomes 0.
The square of NN $t$-matrices for proton-proton and non-charge exchange 
neutron-proton scattering decreases monotonically as a function of the 
momentum transfer.
Thus the contribution from the $E$-term becomes smaller than that from 
the $F$-term through the integration of the momentum transfers.

The results for $^{40}$Ca at 494 and 346 MeV are given in 
Figs.~11 and 12, respectively. 
Since $^{40}$Ca is a spin-saturated nucleus, 
the relation (\ref{eq:spnrsp}) becomes a better approximation.
We guess that the results are more reliable than those for $^{12}$C.
The two-step contributions in comparison with the one-step contributions 
are larger than those for the $^{12}$C case.
The momentum transfer $q^{\rm int}$ for the reactions at 494 MeV is 
1.655~(1/fm) ($q_{\rm lab}$ is 1.69--1.71~(1/fm)).
For the $(p,p')$ scattering and the spin longitudinal and spin 
transverse $^{40}$Ca$(p,n)$ reactions, the ratios are 7.3$\%$, 45$\%$ 
and 87$\%$ at $\omega_{\rm lab}=120$ MeV, respectively.

At 346 MeV the momentum transfer $q^{\rm int}$ is 1.625~(1/fm)
($q_{\rm lab}$ is 1.66--1.68~(1/fm)).
The ratios reach 13$\%$ for the $(p,p')$ scattering, 49$\%$ 
for the spin longitudinal $(p,n)$ reaction and 88$\%$ for the spin 
transverse reaction at $\omega_{\rm lab}=120$ MeV.
We can say that the contribution from the two-step processes is larger 
at 346 MeV than at 494 MeV only for the $(p,p')$ scattering in the 
$^{40}$Ca case. Here we have confirmed again that 
the two-step contribution to the spin transverse reaction is much larger 
than that to the spin longitudinal reaction.

\section{Summary and conclusion}\label{conclusion}

We constructed a formalism for the two-step direct reaction within the 
framework of the plane wave approximation in order to compare its 
contribution with the one-step process. 
With this approximation we factored the nucleon-nucleus $T$-matrix 
into the NN $t$-matrix and the transition density in the one-step 
formalism. 
In the two-step formalism we expressed the two-step cross section with 
a nondiagonal response function with respect to the momentum transfer 
${\bf q}$. 

In the Green's function we removed the effect of absorption, because the 
motion of the projectile and the ejectile was described with plane 
waves. 
We applied the on-energy shell approximation to the Green's function. 

With this formalism, we calculated the cross sections of the $^{12}$C,
$^{40}$Ca$(p,p')$ scattering and the spin longitudinal and spin 
transverse cross sections of the $^{12}$C,$^{40}$Ca$(p,n)$ reactions 
at 346 and 494 MeV. 
The scattering angles were set to 22$^\circ$ and 18$^\circ$, 
respectively, for comparison with the experiments 
at LAMPF\cite{Taddeucci} and RCNP.\cite{Wakasa} 

In $(p,p')$ scattering, the contributions from the two-step 
processes were found to be small.
However, there are appreciable contributions from two-step processes in 
the $(p,n)$ reactions. 
In the spin longitudinal $(p,n)$ reaction, the two-step contribution 
is about 30$\%$--50$\%$ of the one-step cross section, 
while the ratio for the spin transverse $(p,n)$ cross sections become 
about 70$\%$--90$\%$ around $\omega_{\rm lab}=120$ MeV. 
We believe that the difference between the two-step contributions for 
the spin longitudinal cross section and the spin transverse cross 
section is due to the difference between the momentun transfer 
dependences of the $E$-term and the $F$-term in the NN scattering 
amplitude.

We found that the contributions of the two-step processes become larger 
as the energy transfer increases in the quasi-elastic region.
We also found that the two-step contributions in comparison with the 
one-step contributions are larger in the $^{40}$Ca case than in the 
$^{12}$C case.

It has been reported that in DWIA calculations the cross sections of the 
spin longitudinal $(p,n)$ reactions are underestimated beyond the 
quasi-elastic peak region, and for the spin transverse $(p,n)$ 
reactions, the calculations amount to only half of the experimental 
results.
We give the DWIA results multiplied by the ratios of the sum of the one- 
and the two-step cross sections to the one-step cross section for 
$^{12}$C in Fig.~13 and for $^{40}$Ca in Fig.~14.
From these figures we can clearly see that the theoretical results 
including the two-step contribution are closer to the experimental 
results than the DWIA results, particularly for $ID_q$ at 494 MeV.

However, our results are still insufficient for explaining all of the 
discrepancy. 
In those regions where the cross sections are still underestimated, 
the response functions that include 2p-2h correlations may have 
some effect.
Further multi-step processes must be investigated in these regions.
A two-step calculation with full distortion is of course needed to 
obtain a more quantitatively reliable conclusion.

\section*{Acknowledgements}

The authors would like to thank K.~Nishida and K.~Kawahigashi for 
providing their FORTRAN program for the response functions.

\begin{flushleft}
{\Large {\bf Appendix}}
\end{flushleft}
\setcounter{equation}{0}
\renewcommand{\theequation}
  {A.\arabic{equation}}

We used the method of Love and Franey \cite{Love} to calculate the NN 
$t$-matrix.
It is written in the form
\begin{eqnarray}
t_{\rm NN}(K_{{\rm lab}},q)&=&
    [\tilde{V}_{\rm SO}^{\rm C}(q)-
    \tilde{V}_{\rm SO}^{\rm C}(Q)]P_{\rm SO}
   +[\tilde{V}_{\rm SE}^{\rm C}(q)+
     \tilde{V}_{\rm SE}^{\rm C}(Q)]P_{\rm SE} \nonumber \\
 &+&[\tilde{V}_{\rm TO}^{\rm C}(q)-
     \tilde{V}_{\rm TO}^{\rm C}(Q)]P_{\rm TO}
   +[\tilde{V}_{\rm TE}^{\rm C}(q)+
     \tilde{V}_{\rm TE}^{\rm C}(Q)]P_{\rm TE} \nonumber \\
 &+&\frac{i}{4}[Q\tilde{V}^{\rm LSO}(q)+q\tilde{V}^{\rm LSO}(Q)]
    (\sigma_0+\sigma_i)\cdot\hat{n}P_{\rm TO} \nonumber \\
 &+&\frac{i}{4}[Q\tilde{V}^{\rm LSE}(q)-q\tilde{V}^{\rm LSE}(Q)]
    (\sigma_0+\sigma_i)\cdot\hat{n}P_{\rm TE} \nonumber \\
 &-&[\tilde{V}^{\rm TNO}(q)S_{12}^{\rm ODD}(\hat{q})
    -\tilde{V}^{\rm TNO}(Q)S_{12}^{\rm ODD}(\hat{Q})] \nonumber \\
 &-&[\tilde{V}^{\rm TNE}(q)S_{12}^{\rm EVEN}(\hat{q})
    +\tilde{V}^{\rm TNE}(Q)S_{12}^{\rm EVEN}(\hat{Q})] ,
\end{eqnarray}
where 
\begin{equation}
q=2k_{\rm cm}{\rm sin}(\theta/2),
\end{equation}
\begin{equation}
Q=2k_{\rm cm}{\rm cos}(\theta/2),
\end{equation}
and
\begin{eqnarray}
P_{\rm SO}&=&P_{s=0}P_{t=0}
  =\frac{1-\sigma_0\cdot\sigma_i}{4}\frac{1-\tau_0\cdot\tau_i}{4}, \\
P_{\rm SE}&=&P_{s=0}P_{t=1}
  =\frac{1-\sigma_0\cdot\sigma_i}{4}\frac{3+\tau_0\cdot\tau_i}{4}, \\
P_{\rm TO}&=&P_{s=1}P_{t=1}
  =\frac{3+\sigma_0\cdot\sigma_i}{4}\frac{3+\tau_0\cdot\tau_i}{4}, \\
P_{\rm TE}&=&P_{s=1}P_{t=0}
  =\frac{3+\sigma_0\cdot\sigma_i}{4}\frac{1-\tau_0\cdot\tau_i}{4}, \\
S_{12}^{\rm ODD}(\hat{q})&=&S_{12}(\hat{q})P_{t=1}
  =(3\sigma_0\cdot\hat{q}\sigma_i\cdot\hat{q}-\sigma_0\cdot\sigma_i)
   \frac{3+\tau_0\cdot\tau_i}{4}, \\
S_{12}^{\rm EVEN}(\hat{q})&=&S_{12}(\hat{q})P_{t=0}
  =(3\sigma_0\cdot\hat{q}\sigma_i\cdot\hat{q}-\sigma_0\cdot\sigma_i)
   \frac{1-\tau_0\cdot\tau_i}{4} ,
\end{eqnarray}
with the projectile wave number $k_{\rm cm}$ in the NN c.m. system.
The potential depths are given as
\begin{eqnarray}
\tilde{V}_{s\pi}^{\rm C}(k)&=&
  4\pi\sum_i\frac{V_{s\pi ,i}^{\rm C}(R_i^{\rm C})^3}
  {1+(kR_i^{\rm C})^2}, \\
\tilde{V}^{{\rm LS}\pi}(k)&=&
  8\pi\sum_i\frac{V_i^{{\rm LS}\pi}k(R_i^{\rm LS})^5}
  {[1+(kR_i^{\rm LS})^2]^2}, \\
\tilde{V}^{{\rm TN}\pi}(k)&=&
  32\pi\sum_i\frac{V_i^{{\rm TN}\pi}k^2(R_i^{\rm TN})^7}
  {[1+(kR_i^{\rm TN})^2]^3},
\end{eqnarray}
where $k=q$ or $Q$, $s$ is the spin singlet~(S) or spin triplet~(T),
and $\pi$ is the parity in the NN system.
The summations over $i$ are taken over several ranges.
It is these values, $V_{s\pi ,i}^{\rm C},V_i^{{\rm LS}\pi}$
and $V_i^{{\rm TN}\pi}$ that are given in the table in 
Ref.~\cite{Franey}.

Defining
\begin{eqnarray}
  t_{\rm SO}^{\rm C}&\equiv&\tilde{V}_{\rm SO}^{\rm C}(q)-
                            \tilde{V}_{\rm SO}^{\rm C}(Q), \nonumber \\
  t_{\rm SE}^{\rm C}&\equiv&\tilde{V}_{\rm SE}^{\rm C}(q)+
                            \tilde{V}_{\rm SE}^{\rm C}(Q), \nonumber \\
  t_{\rm TO}^{\rm C}&\equiv&\tilde{V}_{\rm TO}^{\rm C}(q)-
                            \tilde{V}_{\rm TO}^{\rm C}(Q), \nonumber \\
  t_{\rm TE}^{\rm C}&\equiv&\tilde{V}_{\rm TE}^{\rm C}(q)+
                            \tilde{V}_{\rm TE}^{\rm C}(Q), \nonumber \\
  t^{\rm LSO}&\equiv&\frac{1}{4}[Q\tilde{V}^{\rm LSO}(q)+
                                 q\tilde{V}^{\rm LSO}(Q)], \nonumber \\
  t^{\rm LSE}&\equiv&\frac{1}{4}[Q\tilde{V}^{\rm LSE}(q)-
                                q\tilde{V}^{\rm LSE}(Q)], \nonumber \\
  t^{\rm TNO}&\equiv&\tilde{V}^{\rm TNO}(q),\tilde{V}^{\rm TNO}(Q), 
   \nonumber \\
  t^{\rm TNE}&\equiv&\tilde{V}^{\rm TNE}(q),\tilde{V}^{\rm TNE}(Q) ,
\end{eqnarray}
and noting Eqs.~(\ref{eq:sixterm}) and (\ref{eq:ATOT}), 
one obtains the relations
\begin{eqnarray}
\eta A_0&=&\frac{1}{16}( t_{\rm SO}^{\rm C}+3t_{\rm SE}^{\rm C}+
                        9t_{\rm TO}^{\rm C}+3t_{\rm TE}^{\rm C}), \\
\eta A_1&=&\frac{1}{16}(-t_{\rm SO}^{\rm C}+ t_{\rm SE}^{\rm C}+
                        3t_{\rm TO}^{\rm C}-3t_{\rm TE}^{\rm C}), \\
\eta B_0&=&
    \frac{1}{16}(-t_{\rm SO}^{\rm C}-3t_{\rm SE}^{\rm C}+
                 3t_{\rm TO}^{\rm C}+ t_{\rm TE}^{\rm C}), \nonumber \\
 &+&\frac{1}{4}
  (3t^{\rm TNO}(q)+t^{\rm TNE}(q)-3t^{\rm TNO}(Q)+t^{\rm TNE}(Q)), \\
\eta B_1&=&
    \frac{1}{16}( t_{\rm SO}^{\rm C}- t_{\rm SE}^{\rm C}+
                  t_{\rm TO}^{\rm C}- t_{\rm TE}^{\rm C}) \nonumber \\
 &+&\frac{1}{4}
  ( t^{\rm TNO}(q)-t^{\rm TNE}(q)- t^{\rm TNO}(Q)-t^{\rm TNE}(Q)), \\
\eta C_0&=&\frac{1}{4}(3t^{\rm LSO}+t^{\rm LSE}), \\
\eta C_1&=&\frac{1}{4}( t^{\rm LSO}-t^{\rm LSE}), \\
\eta E_0&=&\eta B_0+\frac{1}{4}(-9t^{\rm TNO}(q)-3t^{\rm TNE}(q)), \\
\eta E_1&=&\eta B_1+\frac{1}{4}(-3t^{\rm TNO}(q)+3t^{\rm TNE}(q)), \\
\eta F_0&=&\eta B_0+\frac{1}{4}( 9t^{\rm TNO}(Q)-3t^{\rm TNE}(Q)), \\
\eta F_1&=&\eta B_1+\frac{1}{4}( 3t^{\rm TNO}(Q)+3t^{\rm TNE}(Q)) .
\end{eqnarray}

The corresponding $t$-matrix in the nucleon-nucleus system can, to a 
good approximation, be written by replacing $Q$ with the wave number of 
the projectile in the nucleon-nucleus system $k_{\rm i}$, which is 
expressed as 
\begin{equation}
k_{\rm i}^2 \equiv
   M^2A\beta\left[\frac{1+\alpha}{1+\beta}\right], \hspace{5mm}
\alpha\equiv\frac{K_{\rm lab}}{2M}, \hspace{5mm}
\beta\equiv\frac{4\alpha A}{(A+1)^2}, 
\end{equation}
where $A$ is the mass number of the target nucleus. 

Moreover, for calculating nucleon-nucleus scattering, an $A$ dependent 
kinematic modification is required. 
This modification is provided by the transformation of the $t$-matrices 
given by
\begin{equation}
t_{\rm NA}=\frac{\epsilon_{\rm cm}^2}{\epsilon_{\rm i}\epsilon_{\rm t}}
             t_{\rm NN}, 
\end{equation}
\begin{equation}
\epsilon_{\rm cm}^2 = M^2(1+\alpha), \hspace{5mm}
\epsilon_{\rm i}^2 = M^2+k_{\rm i}^2, \hspace{5mm}
\epsilon_{\rm t}^2 = M^2+\left(\frac{k_{\rm i}}{A}\right)^2,
\end{equation}
where $\epsilon_{\rm i}\ (\epsilon_{\rm t})$ is the total energy of the 
incident~(target) nucleon in 
the nucleon-nucleus system, 
and $\epsilon_{\rm cm}$ is the total energy of the incident nucleon in 
the NN system.

\end{document}